\documentclass[prb,aps]{revtex4}

\newcommand{\beq}{\begin{eqnarray}}
\newcommand{\eeq}{\end{eqnarray}}
\renewcommand{\vec}[1]{{\mathbf{#1}}}
\begin{document}
\draft

\title
{Absence of Phase Stiffness in the Quantum Rotor Phase Glass}

\author{ Philip Phillips}
\affiliation{Loomis Laboratory of Physics,
University of Illinois at Urbana-Champaign,
1100 W.Green St., Urbana, IL., 61801-3080}
\author{Denis Dalidovich}
\affiliation{National High Field Magnetic Laboratory,Florida State University,
Tallahassee, Florida 32310}
\vspace{.05in}

\begin{abstract}
We analyze here the consequence of local rotational-symmetry breaking in the 
quantum spin (or phase) glass state of the quantum random rotor model.
By coupling the spin glass order parameter directly to a vector potential, 
we are able to compute whether the system is resilient (that is, possesses 
a phase stiffness) to a uniform rotation in the presence of random 
anisotropy.  We show explicitly that the O(2) vector spin glass has 
no electromagnetic response indicative of a superconductor at mean-field 
and beyond, suggesting the absence of phase stiffness.
This result confirms our earlier finding (PRL, {\bf 89}, 27001 (2002))
that the phase glass is metallic, due to the main contribution to 
the conductivity arising from fluctuations of the superconducting
order parameter.  In addition, our finding that the spin stiffness vanishes
in the quantum rotor glass is consistent with the absence of a transverse
stiffness in the Heisenberg spin glass found by Feigelman and Tsvelik (Sov.
Phys. JETP, {\bf 50}, 1222 (1979).  
\end{abstract}

\maketitle

\section{Introduction}

Spin glasses are characterized by the freezing of local spins along 
random non-collinear directions.  Because each spin points in a preferred 
direction, locally spin rotational symmetry is broken.  Nonetheless, globally
rotational symmetry is preserved because spin glasses have no
net magnetization.  We consider here the $O(2)$ quantum rotor model where
the exchange interactions are random.  As this model
is isotropic in rotor space, a global rotation of all of the rotors is an
exact symmetry, even in the glass phase.  Nonetheless, in the glass state, 
a global rotation of all of the spins around any axis generates a new 
state which is distinguishable from the
original unrotated state. Because such uniform rotations are generated
by the group $SO(2)$, the spin glass state breaks $SO(2)$ symmetry.  
All such states are energetically degenerate as a result
of the inherent isotropy in rotor space.  As a result of the broken
SO(2) symmetry, it is reasonable to expect that a massless bosonic mode 
should exist. 

In the strict sense, a physical system possesses a non-zero phase 
rigidity if upon a uniform rotation of the phase, the free energy increase 
is of the form, 
\beq\label{eq1}
\Delta F=\frac{\rho_s}{2}\int d^2r|\nabla\theta|^2,
\eeq
where $\rho_s$ is the spin or superfluid stiffness and $\theta$ is the 
collective phase variable. Consequently a spin-wave
mode with a dispersion $\omega=\pm ck$ would be an experimental signature 
of a spin stiffness consistent with Eq. (\ref{eq1}).  Experimentally, however,
no such mode has ever been found in either neutron scattering or thermal
measurements\cite{expts1,expts2,expts3,expts4} on spin glasses.  This 
failure might be attributed to that fact that over-damped modes
and/or low energy excitations conspire to make $\rho_s$ undetectable.  
Theoretically, in the  phenomenological hydrodynamic account, 
Halperin and Saslow\cite{hs} assumed that $\rho_s\ne 0$. They did caution 
the reader that the existence 
of a stiffness in a spin glass is subtle and, in all likelihood, 
doubtful as a result of the preponderance of experimental evidence for a 
large density of low-energy
excitations that could over-damp the spin-wave mode.
This conclusion is supported by extensive numerical simulations
by Walker and Walstedt\cite{ww} who found no evidence for the characteristic
$\omega^2$ vanishing of the low-energy modes.  
Two microscopic calculations of the spin stiffness exist.  Feigel'man and Tsvelik\cite{ft} developed a real-time diagrammatic technique for the Heisenberg spin glass and showed
explicitly that the spin stiffness vanishes. This result is particularly robust because it follows from a simple permutation symmetry
of the spin correlators\cite{ft}.   
 Within the replica formalism
of a Heisenberg spin glass
Kotliar, Sompolinsky, and Zippelius\cite{kotliar1,ksz} formulated a 
mean-field description of the {\bf single-valley} stiffness constant.
This limit is relevant at sufficiently short times that the spin glass
remains trapped in a single configuration.  In this limit, the stiffness constant is non-zero\cite{kotliar1,ksz}. However, in the full statistical mechanical 
treatment of the problem in which hopping among the myriads of 
valleys in the energy landscape of a spin glass are allowed, the stiffness vanishes\cite{kotliar1,ksz,by}.  This result
 implies that the spin stiffness is a transient effect approaching zero in
the equilibrium or long-time limit. In this limit, a new massless mode
dispersing as $k^4$ emerges which leads to the vanishing of the spin stiffness,
as in the real-time formalism\cite{ft}.  Hence, there is a consilience between the replica and real-time formalisms that the stiffness constant vanishes in the Heisenberg spin glass.

For quantum spin glasses, no calculation of the stiffness exists. Nontheless,
we expect the same physics to be valid.  Namely, as long as the system
can relax and hop among all of the configurations of the spin glass, 
the stiffness should vanish.  For example, in quantum spin glasses, 
quantum tunneling among the various local minima in the spin glass 
landscape is permitted, thereby leading to 
a vanishing of the stiffness. 
This problem is particularly current because we have recently 
proposed\cite{dppg} that the bosonic excitations arising from  
fluctuations of the superconducting order parameter in the
glassy phase, lead to a metallic conductivity at zero temperature. 
In the Gaussian approximation, this conductivity
$\sigma_{bos}$ diverges as $1/m^4$ upon approaching
the superconducting phase
($m$ is the inverse correlation length of the superconducting fluctuations).
A free energy density of the form of Eq. ({\ref{eq1}), however, leads to a 
superconducting response. Hence,
should the phase glass itself have a well-defined stiffness, 
then the bosonic conductivity, though intriguing, would be irrelevant 
as it would be dwarfed by the infinite conductivity arising from 
the excitations related to the glassy order parameter.
We show here explicitly that this is not the case, at least at the mean-field
level. Rather than attempting to calculate the phase stiffness from the 
free energy, we consider the linear response regime and couple the spin glass 
order parameter to the appropriate vector potential. 
Second, we compute the role of replica symmetry breaking (RSB) on the 
bosonic contribution to the conductivity.  We show that weak RSB
does not affect the metallic character of the conductivity 
as $T\rightarrow 0$.
Consequently, the Bose metallic phase found earlier\cite{dppg} is robust 
and constitutes the only known example of a metallic phase in 2D in 
the presence of disorder.

\section{Phase Stiffness}

The starting point for our analysis is the $O(2)$ quantum rotor model,
\beq
H=-E_C\sum_i\left(\frac{\partial}{\partial\theta_i}\right)^2-
\sum_{\langle i,j\rangle} J_{ij}\cos(\theta_i-\theta_j-A_{ij}), 
\eeq
where $A_{ij}=(e^* /\hbar) \int_i^j {\vec A}\cdot d{\vec l}$ ($e^* =2e$).
The Josephson couplings are assumed to be random and governed by a 
distribution
\beq P(J_{ij})=\frac{1}{\sqrt{2\pi
J^2}}\exp{\left[-\frac{(J_{ij}-J_0)^2}{2J^2}\right]} \eeq 
with non-zero mean, $J_0$ and $J$ the variance. When the distribution has 
a non-zero mean, three phases are possible: 1) disordered paramagnet, 
2) quantum phase glass, and 3) superconductor.  Because the existence of 
the spin 
stiffness in the spin glass can be answered with the simpler model 
with zero mean ($J_0=0$), we utilize this model at the outset.  For a 
random system, the technique for treating disorder is now standard: 
1) replicate the partition function,
2) perform the average over disorder and 3) introduce the appropriate 
fields to decouple the interacting terms that arise.  As the corresponding 
action has been detailed previously\cite{dp2,rsy}, we will provide 
additional steps that are necessary to determine how the electromagnetic 
gauge couples to the spin glass order parameter. We write the replicated 
partition function as
\beq\label{zn}
\overline{Z^n}=\int {\cal D}\theta_i{\cal D}J_{ij}e^{-S}
\eeq
where the Euclidean action is given by
\beq
S=\int_0^\beta d\tau\left\{\sum_{i,a}\frac{1}{4E_C}\left(\frac{\partial
\theta_i^a (\tau)}{\partial\tau}\right)^2  -\sum_a\sum_{\langle ij\rangle}
J_{ij}\cos\left[\theta_i^a (\tau) -\theta_j^a (\tau) - 
A_{ij}(\tau)\right]\right\},
\eeq
where the superscript $a$ represents the replica index.
For $J_0=0$, the integration over $J_{ij}$
 in Eq. (\ref{zn}) results in the effective action,
\beq
S_{\rm  eff}&=&\int_0^\beta d\tau\sum_{i,a}\frac{1}{4E_C}\left(
\frac{\partial\theta_i^a}{\partial\tau}\right)^2\nonumber\\
&&+\frac{J^2}{2}\sum_{\rm a,b}\sum_{\langle ij\rangle}
\int_0^\beta\int_0^\beta d\tau d\tau'\frac{1}{4}\sum_{\alpha=+1,-1}
\exp\left\{i\left[\theta_i^{a}(\tau)-\alpha\theta_i^b(\tau')-
\left(\theta_j^{a}(\tau)-\alpha\theta_j^b(\tau')\right)-\left(A_{ij}(\tau)-
\alpha A_{ij}(\tau')\right)\right]\right\}\nonumber\\
&& +c.c.
\eeq
with $\alpha=+1,-1$.  As a result of the sum over $\alpha$, we see that the 
vector potential enters both symmetrically and anti-symmetrically.
To simplify the notation, we introduce the two-component vector
\beq
S^a(\tau)=\left(\cos\theta^a(\tau),\sin\theta^a(\tau)\right) 
\eeq
and the corresponding auxiliary field,
\beq
Q_{\rm \mu\nu}^{ab}(\tau,\tau')=\langle S^a_\mu(\tau) S^b_\nu(\tau')\rangle
\eeq
which will be used in decoupling the action and ultimately determines 
the Edwards-Anderson order parameter for the quantum spin glass transition.  
The remaining steps involve performing the cumulant expansion
and taking the continuum limit.  The final action can be separated into 
the local and gradient parts:
\beq
S_{\rm eff}=S_{\rm loc} +S_{\rm gr}
\eeq
where the local part
\begin{eqnarray} \label{action}
S_{\rm loc}& = & \int d^dx\left\{\frac{1}{\kappa}\int
d\tau\sum_a
\left(r+\frac{\partial}{\partial\tau_1}\frac{\partial}{\partial\tau_2}
\right)Q_{\mu\mu}^{aa}(\vec x ,\tau_1,\tau_2)|_{\tau_1=\tau_2=\tau}
\right.\nonumber\\
&&-\frac{\kappa}{3}\int d\tau_1 d\tau_2
d\tau_3\sum_{a,b,c}Q^{ab}_{\mu\nu}(\vec x ,\tau_1,\tau_2)
Q^{bc}_{\nu\rho}(\vec x ,\tau_2,\tau_3)
Q^{ca}_{\rho\mu}(\vec x ,\tau_3,\tau_1)\nonumber\\
&&\left.+\frac12\int
d\tau\sum_a\left[uQ^{aa}_{\mu\nu}(\vec x ,\tau,\tau)
Q^{aa}_{\mu\nu}(\vec x ,\tau,\tau)
+vQ^{aa}_{\mu\mu}(\vec x ,\tau,\tau)Q^{aa}_{\nu\nu}(\vec x ,\tau,\tau)
\right]\right\}\nonumber\\
&&-\frac{y_1}{6t}\int d^d x\int
d\tau_1d\tau_2\sum_{a,b}\left[Q^{ab}_{\mu\nu}(\vec x ,\tau_1,\tau_2)
\right]^4.
\end{eqnarray}
is identical to that derived previously by Read, Sachdev and Ye\cite{rsy}
and the gradient part
\beq
S_{\rm gr}&=&\int d^dx\int_0^\beta d\tau_1 d\tau_2\sum_{a,b}
\left|\left(\nabla-\frac{ie^*}{\hbar}
\vec A(\vec x,\tau_1)+\frac{ie^*}{\hbar}\vec A(\vec x ,\tau_2)\right)
Q^{ab}_{-}(\vec x ,\tau_1,\tau_2)\right |^2\nonumber\\
&&+\int d^dx\int_0^\beta d\tau_1 d\tau_2\sum_{a,b}
\left| \left(\nabla-\frac{ie^*}{\hbar}
\vec A(\vec x,\tau_1)-\frac{ie^*}{\hbar}\vec A(\vec x ,\tau_2)\right)
Q^{ab}_{+}(\vec x ,\tau_1,\tau_2) \right|^2
\eeq
in which  the vector potential couples both symmetrically and asymmetrically
to combinations of the $Q-$matrices of the same parity.  
Using the fact that $Q^{ab}_{\pm}(\tau_1,\tau_2) \sim
\langle \exp\left[ i(\theta_i^a(\tau)
\pm\theta_i^b(\tau'))\right]\rangle$,
the parity combinations of the $Q-$matrices are defined as follows:
\beq\label{qlin}
Q^{ab}_{\pm}(\vec x , \tau_1,\tau_2)=
\frac12\left[ Q_{11}^{ab}(\vec x ,\tau_1,\tau_2)\mp 
Q_{22}^{ab}(\vec x ,\tau_1,\tau_2)\right]
+\frac{i}{2}\left[Q_{12}^{ab}(\vec x ,\tau_1,\tau_2)\pm 
Q^{ab}_{21}(\vec x ,\tau_1,\tau_2)\right].
\eeq
It is evident that the vector potential enters in a non-time 
translationally invariant manner.  This is a direct consequence of the 
fact that the $Q-$matrices
themselves are a function of two independent times, not simply the difference
of $\tau_1-\tau_2$.
  
To calculate the conductivity, we need to focus entirely on the gradient 
part of the action as this is the only part that couples to 
the vector potential.   
The standard Kubo formula for the spin-glass contribution to the 
longitudinal conductivity takes the form,
\beq
\sigma (i\omega_n)=-\frac{\hbar}{\omega_n}
\lim_{n\rightarrow 0}\frac{1}{n}\int d^d(\vec x-\vec x')\int_{0}^{\beta}
d(\tau-\tau')
\frac{\delta^2\overline{Z^n}}{\delta A_x(\vec x,\tau)
\delta A_x(\vec x',\tau')}
e^{i\omega_n(\tau-\tau')}
\eeq
where we have chosen to orient the vector potential along the x-axis.  
A bit lengthy variational procedure leads to the following result:
\beq\label{sigma1}
\sigma(i\omega_n)&=&\frac{(e^*)^2}{\hbar\omega_n}\frac{1}{n}\sum_{a,b}
\int_{0}^{\beta} d(\tau-\tau')
e^{i\omega_n(\tau-\tau')} \left\{ 4 \int_{0}^{\beta} d\tau_2
\left(\langle |Q^{ab}_{-}(\vec x,\tau,\tau_2)|^2\rangle+
\langle|Q^{ab}_{+}(\vec x,\tau,\tau_2)|^2\rangle\right)
\delta(\tau-\tau')\right.\nonumber\\
&&\left.+4\left(\langle |Q^{ab}_{+}(\vec x,\tau,\tau')|^2\rangle-
\langle|Q^{ab}_{-}(\vec x,\tau,\tau')|^2\rangle\right)
-\int d^d (\vec x-\vec x') \langle J_x (\vec x, \tau)
J_x (\vec x', \tau')\rangle \right\},
\eeq
where the current $ J_x (\vec x, \tau)$ is defined as
\beq\label{curr}
\vec J(\vec x,\tau)&=&\frac{i e^*}{\hbar}\sum_{ab}\sum_{\alpha=+,-}
\int_{0}^{\beta} d\tau_1\left[
Q^{ab}_{\alpha}(\vec x,\tau,\tau_1)\nabla \left(Q^{ab}_{\alpha}
(\vec x,\tau,\tau_1) \right)^* -{\rm c.c} \right].
\eeq
In deriving this expression for the current, we considered the 
relations 
$Q^{ab}_{+}(\vec x,\tau_2,\tau_1) = Q^{ab}_{+}(\vec x,\tau_1,\tau_2)$ and
$Q^{ab}_{-}(\vec x,\tau_2,\tau_1)= (Q^{ab}_{-}(\vec x,\tau_1,\tau_2))^*$,
that follow from the definition, Eq. (\ref{qlin}).
To evaluate the correlation functions in Eq. (\ref{sigma1}), we need to use 
the Fourier components of the $Q$-fields:
\beq
Q_{\mu\nu}^{ab}(\vec x,\tau_1,\tau_2)&=&\int\frac{d^d\vec k}{(2\pi)^d}
\frac{1}{\beta^2}\sum_{\omega_1,\omega_2} Q_{\mu\nu}^{ab}
(\vec k_,\omega_1,\omega_2)
e^{-i(\vec k\cdot\vec x-\omega_1\tau_1-\omega_2\tau_2)},
\eeq
and take into account the relations between $Q_{\pm}^{ab}$ and 
$Q_{\mu\nu}^{ab}$ given by Eq. (\ref{qlin}). 
The general ansatz for the Fourier transformed $Q$-matrices,
\beq\label{sg1}
Q^{ab}_{\mu\nu}(\vec k,\omega_1,\omega_2)= \beta (2\pi)^d\delta^d(\vec k)
\delta_{\mu\nu}[\beta q^{ab}\delta_{\omega_1,0}\delta_{\omega_2,0}+
\delta^{ab}\delta_{\omega_1+\omega_2,0}D(\omega_1)]+
\tilde{Q}^{ab}_{\mu\nu}(\vec k,\omega_1,\omega_2)
\eeq 
consists of the spatially uniform mean-field part and the 
fluctuating spatial component, $\tilde{Q}^{ab}$.  In Eq. (\ref{sg1}),
\beq\label{D}
D(\omega)=-|\omega|/\kappa, 
\eeq
while the off-diagonal elements 
of $q^{ab}$ constitute the ultrametric Parisi matrix \cite{rsy}
\begin{eqnarray}
q(s)=\left\{\begin{array}{ll} (s/s_1) q_{EA}&0<s<s_1,\nonumber\\
q_{EA}&s_1<s<1,
\end{array}
\right.
\end{eqnarray}
in which $s_1=2 y_1 q_{EA}T/\kappa$, and $q_{EA}$ is the Edwards-Anderson 
order parameter ($q^{aa}=q_{EA}$).

We substitute then this ansatz into Eq. (\ref{sigma1})  
and obtain that $\sigma(i\omega_n)$ consists of three parts,
\beq
\sigma(i\omega_n)=\sigma^{(1)}(i\omega_n)+\sigma^{(2)}(i\omega_n)
+\sigma^{(3)}(i\omega_n).
\eeq
$\sigma^{(1)}(i\omega_n)$ is given by
\beq\label{sig1}
\sigma^{(1)}(i\omega_n)=\frac{16e^2}{\hbar \omega_n}\frac{4 q_{EA}\Delta_q}{3}
\Pi(i\omega_n)
\eeq
where 
\beq\label{Pi}
\Pi(i\omega_n)=\int_0^\beta
d\tau e^{i\omega_n\tau}\left[\beta\delta(\tau)-1\right],
\eeq
In the derivation above, we used the result,
$(1/n)\sum_{a,b} q^{ab} q^{ab}=(4/3)q_{EA}\Delta_q$, where
$\Delta_q=q_{EA}-\int_0^1 q(s)ds=q_{EA}s_1 /2$ is the broken ergodicity 
parameter, that vanishes linearly with temperature.  Note, had we assumed 
that the vector potential entered in a time-translationally invariant manner, 
the factor of $-1$ in Eq. (\ref{Pi}) would not be present.  As a result, the 
conductivity would diverge at $\omega_n=0$ as in a superconductor. 
In $\sigma^{(2)}(i\omega_n)$ we collect the terms that contain $D(\omega_n)$:
\beq\label{sig2}
\sigma^{(2)}(i\omega_n)=\frac{16e^2}{\hbar \omega_n}
\left( T\sum_{\omega_m}D^2(\omega_m)- T\sum_{\omega_m}D(\omega_m)
D(\omega_m+\omega_n) -2q_{EA}D(\omega_n) \right).
\eeq

The remaining term, $\sigma^{(3)}(i\omega_n)$ arises from the 
spatially-dependent part 
$\tilde{Q}^{ab}_{\mu\nu}(\vec k,\omega_1,\omega_2)$ of the $Q$-matrices.
Writing the expression for the current, Eq. (\ref{curr}), in
two parts,
\beq\label{curr1}
{\vec J}_1 (\vec x,\tau)=-\frac{2e^{*}}{\hbar}\sum_{ab} 
\int\frac{d^d k}{(2\pi)^d} \frac{1}{\beta^2}\sum_{\omega_1,\omega_2}
\vec k \left( \beta q^{ab}\delta_{\omega_2,0}+
\frac{1}{\beta}\sum_{\omega_2}D(\omega_2) \right)
\left[ \tilde{Q}_{-}^{ab}(\vec k,\omega_1,\omega_2)
e^{i(\vec k\cdot\vec x-\omega_1\tau -\omega_2\tau)} +c.c \right],
\eeq
\beq\label{curr2}
{\vec J}_2 (\vec x,\tau)=-\frac{2e^{*}}{\hbar}\sum_{ab}\beta
\sum_{\alpha=\pm} 
\int\frac{d^d k_1}{(2\pi)^d} \frac{d^d k_2}{(2\pi)^d}
\frac{1}{\beta^3}\sum_{\omega_1,\omega_2,\omega_3}
(\vec k_1 +\vec k_2) \tilde{Q}_{\alpha}^{ab}(\vec k_1,\omega_1,\omega_3)
\left( \tilde{Q}_{\alpha}^{ab}(\vec k_2,\omega_2,\omega_3)\right)^*
e^{i(\vec k_2 -\vec k_1)\cdot\vec x}
e^{i(\omega_1-\omega_2)\tau},
\eeq
we observe that the contribution from ${\vec J}_1 (\vec x,\tau)$ vanishes as
a result of integration over $d^d (\vec x -\vec x')$. The remaining part 
leads to the result that
\beq\label{sig3}
\sigma^{(3)}(i\omega_n)&=&\frac{4(e^*)^2}{\hbar\omega_n}
\frac{\beta}{n}\sum_{a,b}\sum_{\alpha=+,-}
\int \frac{d^d k}{(2\pi)^d}\frac{1}{\beta^2}\sum_{\omega_1,\omega_2} 
\left[ {\cal G}^{ab}_{\alpha}(\vec k,\omega_1,\omega_2)\right.\nonumber\\
&&\left.-4 k_x^2 
\Gamma^{ab}_{\alpha}(\vec k, \omega_1,\omega_2;\omega_n)
{\cal G}^{ab}_{\alpha}(\vec k, \omega_1,\omega_2)
{\cal G}^{ab}_{\alpha}(\vec k, \omega_1,\omega_2+\omega_n) \right].
\eeq 
In Eq. (\ref{sig3})
\beq
{\cal G}^{ab}_{\pm}(\vec k, \omega_1,\omega_2)=
\langle {\tilde Q}_{\pm}^{ab}(\vec k,\omega_1,\omega_2)
{\tilde Q}_{\pm}^{ab}(-\vec k,-\omega_1,-\omega_2)\rangle=
\frac14 \sum_{\mu,\nu=1,2}{\cal G}^{ab}_{\mu\nu}(\vec k, \omega_1,\omega_2),
\eeq
is the exact propagator for the fluctuations of the ${\tilde Q}$-fields.
The first term is the diamagnetic contribution, while the second is 
paramagnetic and can be formally represented by the standard bubble 
diagrams\cite{sachbook} and
$\Gamma^{ab}_{\alpha}(\vec k, \omega_1,\omega_2;\omega_n)$ is 
the corresponding vertex function.
 
We discuss first the contribution $\sigma^{(1)}(\omega_n)$. 
The explicit frequency dependence of this part is given simply by the 
prefactor $\Pi(i\omega_n)/\omega_n$. Should a phase stiffness exist, 
this prefactor would be simply proportional to 
$1/\omega_n$, which when analytically continued would yield the 
standard electromagnetic response for the conductivity of a 
superconductor.  However, this is not 
the case here.  The integral in Eq. (\ref{Pi}) is simply 
$\beta (1-\delta_{\omega_n,0})$ effectively removing thus the 
divergence at zero frequency, unlike what would be the case had we assumed 
that the vector potential entered the action in a time-translationally 
invariant manner.  Note that such an expression although
not analytic at $\omega_n=0$ does not violate causality because
it is, nonetheless, analytic in either the upper or lower half planes.
Hence, the $O(2)$ quantum phase glass has a vanishing stiffness in the 
limit $\omega_n=0$, which of course is the physically relevant regime for the 
dc conductivity. It is in this limit that explorations of all available 
minima are possible.  

To see this result more systematically, we analytically continue 
$\Pi(i\omega_n)$ using a Hilbert transformation.  The denominator of 
Eq. (\ref{sig1}) can be analytically continued trivially, 
$i\omega_n\rightarrow \omega+i\eta$, where $\eta$ is a positive
infinitesimal.
We write the numerator as
\beq\label{piomega}
\Pi(i\omega_n)=\int_0^\beta d\tau e^{i\omega_n\tau}\Pi(\tau),
\quad\Pi(\tau)=\beta\delta(\tau)-1\equiv\Pi_1(\tau)-\Pi_2(\tau)
\eeq
Although $\Pi_1(\tau)$ is not an analytic function, we can construct its 
analytical continuation using the conformal invariance condition, 
$\delta(\tau)=\delta(\tau+\beta)$. Performing the integration
over the first term in Eq. (\ref{piomega}), we obtain that 
$\Pi_1(\omega)=\beta$. Because
$\Pi_2(\tau)=1$ is an analytic function, we adopt the spectral representation
\beq\label{conf}
\Pi_2(\tau)=\frac{1}{\pi}\int_{-\infty}^{\infty}\frac{e^{-\tau\epsilon}
\Pi_2^{''}(\epsilon) d\epsilon}{1-e^{-\beta\epsilon}}
\eeq
valid for Bose systems,
where $\Pi_2(\epsilon)=\Pi_2^{'}(\epsilon)+i\Pi_2^{''}(\epsilon)$. 
This representation is most convenient for constructing the analytical 
continuation\cite{parcollet}.  Once we know
$\Pi_2^{''}(\epsilon)$, we can obtain both the real and the imaginary 
parts for real frequencies using the Hilbert transformation,
\beq\label{hilbert}
\Pi_2(\omega)=\frac{1}{\pi}\int \frac{d\epsilon \Pi_2^{''}(\epsilon)}
{\epsilon-\omega-i0^+}.
\eeq  
Solving Eq. (\ref{conf}) with $\Pi_2(\tau)=1$ yields
\beq
\Pi^{''}_2(\epsilon)=2\pi\delta(\epsilon)\sinh \frac{\epsilon\beta}{2}
=\pi\beta\epsilon\delta(\epsilon).
\eeq

The real part is determined by the principal value
\beq
\Pi_2^{'}(\omega)={\rm P}\int_{-\infty}^{\infty} 
\frac{\beta \epsilon\delta(\epsilon)d\epsilon}{\epsilon-\omega}=
\beta f(\omega)=\beta \left\{\begin{array}{ll} 1& \omega=0\nonumber\\
0& \omega\ne 0
\end{array}
\right. 
\eeq
Obtained in this fashion, the real and imaginary parts of $\Pi_2(\omega)$ 
formally satisfy the Kramers-Kronig relations. However, both are 
not regular functions. Hence, it is more convenient to treat the real
and imaginary parts of $\Pi_2(\omega)$ as limits of two analytic 
functions.  For example, from the regular function, 
\beq\label{limits}
g(\omega)=\beta\left(
\frac{\eta^2}{\eta^2+\omega^2}+
i\frac{\eta\omega}{\eta^2+\omega^2}\right),
\eeq
whose real and imaginary parts satisfy the Kramers-Kronig
relations, we obtain the correct limit for $\Pi_2(\omega=0)=1$ simply
from $g(\omega=0)=1$, and for $\omega\ne 0$ the limiting procedure,
$\lim_{\eta\rightarrow 0}g(\omega)=\Pi_2(\omega\ne 0)=0$.   As a result,
the limits, $\omega=0, \eta\rightarrow 0$ and $\eta=0,\omega\rightarrow 0$ 
do not
commute, a fact which must be considered when we construct the $\omega=0$ 
conductivity. The correct order of limits is $\eta\rightarrow 0,\omega=0$.
Nonetheless, the advantage of writing
$\Pi_2(\omega)$ in this fashion is that for any non-zero $\eta$, the real 
and imaginary parts of this $g(\omega)$ 
obey the Kramers-Kronig relations. Combining this representation
with $\Pi_1(\omega)=1$ and $i\omega_n\rightarrow \omega+i\delta$, we
obtain the analytically continued form for the frequency dependence 
of the conductivity
\beq
\frac{\Pi(i\omega_n)}{\omega_n}\rightarrow
 \beta\left[ \frac{\omega^2(\eta+\delta)}
{(\eta^2+\omega^2)(\eta^2+\omega^2)} +i\frac{\omega^3 -\eta\delta\omega}
{(\eta^2+\omega^2)(\delta^2+\omega^2)} \right] =\left\{\begin{array}
{ll} 0 &\omega=0,\quad \lim_{\delta\rightarrow 0,\eta\rightarrow 0}\nonumber\\
i\beta/\omega &\omega\ne 0,\quad\lim_{\delta\rightarrow 0,\eta\rightarrow 0}
\end{array}
\right.
\eeq
Recall, the correct $\omega=0$ limit is recovered by setting $\omega=0$ and
then taking the limit, $\eta\rightarrow 0$.
We find then that the contribution of $\sigma^{(1)}(\omega)$ to the 
conductivity is purely imaginary. The absence of the real 
part and, as a result, a formal violation of the Kramers-Kronig
relations here is tied to the presence of the non-analytic function 
$\delta(\tau)$ in Eq. (\ref{Pi}).  Such non-analyticity at $\omega=0$ is 
permissible
because the requirement of causality is analyticity in either the upper
or lower half planes. 

To evaluate the $\omega\rightarrow 0$ limit of $\sigma^{(2)}(\omega )$
we must analytically continue the difference of the first two terms
in Eq. (\ref{sig2}). Using Eq. (\ref{D}) we obtain\cite{otterlo,dp} that
\beq\label{s2}
\sigma^{(2)}(\omega=0)=\frac{16e^2}{\hbar} \left( 
\frac{2q_{EA}}{\kappa} +\frac{2}{\pi\kappa^2}\int_0^{\Lambda_{\omega}}
z\coth\frac{z}{2T} dz \right),
\eeq
and is some regular function of the infrared cutoff $\Lambda_{\omega}$
and temperature. We see that the contribution $\sigma^{(2)}(\omega=0)$
is non-critical and metallic. 

Proceeding to the third term, $\sigma^{(3)}(\omega)$, we
 first 
notice that the exact calculation of the 
propagator ${\cal G}^{ab}_{\mu\nu}(\vec k,\omega_1,\omega_2)$,
based on the action Eq. (\ref{action}) is not possible. 
However, at the quantum critical point in the Gaussian approximation, 
\beq\label{gaussprop}
{\cal G}^{ab}_{\mu\nu}(\vec k,\omega_1,\omega_2)=\frac{1}
{k^2+|\omega_1|+|\omega_2|}\equiv {\cal G}_{0}(\vec k,\omega_1,\omega_2),
\eeq
and hence is indepedent of replica and spatial indices. Substitution of 
this simple replica-symmetric 
propagator into Eq. (\ref{sig3}) leads to the zeroth-order result 
for $\sigma^{(3)}(\omega)$ as a result of the replica summation. Because 
the renormalization group equations for the coefficients in the action,
Eq. (\ref{action}), lead to runaway to strong coupling for $d<d_c=8$, 
it is not possible to analyze the behavior of $\sigma^{(3)}(\omega=0)$ for 
the relevant dimensionalities. However, the structure of Eq. (\ref{sig3}) 
allows us to make the conclusion that the
superconducting contribution of the type $\rho_s \delta (\omega)$
is not expected. This can be proven formally by integrating by parts 
the diamagnetic term
and employing the Ward identity. After the analytical continuation 
$\omega_n \rightarrow -i\omega$, we expand the ensuing
 expression over $\omega$. We obtain that
the zero-frequency conductivity obeys the scaling form
\beq
\sigma^{(3)}(\omega=0)=\frac{e^2}{\hbar} \left( 
\frac{T}{\hbar} \right)^{d-2} F \left( \frac{q_{EA} }{T} \right),
\eeq  
albeit the precise form of the function $F(x)$ and, hence, the 
corresponding temperature dependence can not be determined.
  
We have obtained an important result that there is no real 
contribution to the conductivity proportional to $\rho_s \delta (\omega)$.
The vanishing of the stiffness is tied to the nature of the vector 
potential coupling to the glassy order parameter. The vector potential couples
in a non-time translationally invariant manner to the spin glass 
order parameter.  If, however, the system explores only one of the 
myriad of configurations in the glassy landscape,a stiffness appears 
in agreement with the work of Kotliar et. al.\cite{ksz}. However, certainly 
within a single configuration, the origin of time is irrelevant. But this 
is not the most general case.
Quantum mechanically
tunneling to all minima is permitted.  In this case, the stiffness vanishes 
in agreement with the
result\cite{ft,ksz} on the Heisenberg spin glass that the spin stiffness 
is a transient and hence should vanish once tunneling between all 
minima is present. 
This result is robust and expected to hold beyond the mean-field theory.

\section{Bosonic Conductivity: Replica Symmetry Breaking}

Now we generalize our earlier result for the bosonic conductivity.  
Such a contribution arises only in the case of non-zero mean, $J_0\ne 0$.   
In this case an ordered phase exists which in the $O(2)$ case is 
a superconductor. Hence, in the presence of non-zero mean, a new 
order parameter
\beq
\Psi^a_\mu(\vec k,\tau)=\langle S^a_\mu(\vec k,\tau)\rangle
\eeq
which is determined by the expectation value of the rotor spin.  
On the spin glass side of the phase diagram, the bosonic excitations
of the superconductor develop a mass, $m$ which is equivalent to the 
inverse correlation length for phase coherence. In the presence of 
bosonic excitations, the free energy
contains the additional terms,
\beq\label{fen}
&&{\Delta\cal F}[\Psi,Q]=
\sum_{a,\mu, k,\omega_n}(k^2+\omega_n^2+m^2)
|\Psi_\mu^a(\vec k,\omega_n)|^2 \nonumber\\
&&-\frac{1}{\kappa t}\int d^d x\int d\tau_1
d\tau_2\sum_{a,b,\mu, \nu}\Psi_\mu^a(x,\tau_1)
[\Psi^b_\nu(x,\tau_2)]^{*}
Q_{\mu\nu}^{ab}(x,\tau_1,\tau_2)\nonumber\\
&&+\frac{U}{2}\int d\tau\sum_{a,\mu}\left[\Psi_\mu^a(x,\tau)
(\Psi_\mu^a(x,\tau))^{*}\right]^4
\eeq
At the Gaussian level, with the mean-field spin glass ansatz 
(Eq. (\ref{sg1})), the effective Gaussian propagator 
for the bosonic degrees of freedom has the form: 
\beq\label{fgauss}
{\cal F_{\rm gauss}}&=&\sum_{a,\vec k,\omega_n}
(k^2+\omega_n^2+\eta |\omega_n|+m^2)
|\psi^a(\vec k,\omega_n)|^2 \nonumber\\
&&-\beta q\sum_{a,b,\vec k,\omega_n} \delta_{\omega_n,0}
\psi^a(\vec k,\omega_n)[\psi^b(\vec k,\omega_n)]^{\ast}.
\eeq
As we have pointed out previously, the term proportional to $q^{ab}$ in
$F_{\rm gauss}$ cannot be rewritten as an effective mass term because 
this term explicitly couples $\psi$ fields with different replica indices.  
In the case of replica symmetry, that is, $q^{ab}=q_{0}$ for all $a$ and $b$, 
we have shown that the resultant conductivity is non-zero and given by,
\beq\label{sbos}
\sigma_{\rm bos}(\omega=0,T\rightarrow 0)=
\frac{4}{3} \frac{e^2 \eta q_{0}}{h m^4}
\eeq
which smoothly crosses over to $\sigma=\infty$ in the superconducting 
state ($m=0$).  That the bosonic contribution to the conductivity 
should be non-zero is immediately obvious from the $|\omega|$ term 
in the action.  This term arises entirely due to the glass degrees of 
freedom that naturally provide for dissipation to generate a 
metallic state.  

We now generalize this result to include replica symmetry breaking.
Application of the Kubo formula in this case results in a conductivity
\beq\label{bm1}
\sigma(i\omega_n)&=&\frac{2(e^*)^2}{n\hbar\omega_n}T\sum_{a,b,\omega_m}
\int \frac{d^2k}{(2\pi)^2} 
\left[G^{(0)}_{ab}(\vec k,\omega_m)\delta_{ab}\right.\nonumber\\
&&\left.-2 k_x^2 G_{ab}^{(0)}(\vec k, \omega_m)G_{ab}^{(0)}
(\vec k,\omega_m+\omega_n)\right].
\eeq
that depends entirely on the Gaussian propagator for the $\psi$ fields.  To 
evaluate this quantity, we need to invert Eq.(\ref{fgauss}).  
This calculation is difficult to perform for the general type of RSB.
However, it can be readily done using the rules developed by 
Mezard and Parisi\cite{parisi} for inverting an ultrametric matrix 
having a 1-step RSB:
\begin{eqnarray}
q(s)=\left\{\begin{array}{ll} q_0 \quad\quad s<s_c  \nonumber\\
q_1 \quad\quad s_c<s<1
\end{array}
\right.
\end{eqnarray}
To apply the inversion formula detailed in the Appendix II of Ref. [12], 
it is expedient to make the following definitions:
\beq
g=\frac{1}{k^2+\eta|\omega_n|+m^2},\quad \tilde{g}=\frac{1}
{(k^2+\eta|\omega_n|+\Sigma_m)(k^2+\eta|\omega_n|+m^2)},\quad 
\Sigma_m=m^2+\beta\Sigma_1,\quad
\Sigma_1=s_c(q_1-q_0)  
\eeq
Application of the inversion formula\cite{parisi} results in the diagonal
\beq\label{propdia}
\tilde{G}=g+\beta\delta_{\omega_{n,0}}\Sigma_1\tilde{g}
\frac{1-s_c}{s_c}+\beta q_0\delta_{\omega_{n,0}} g^2 
\eeq
and the off-diagonal elements
\beq\label{propoffdia}
G(s)=\beta q_0\delta_{\omega_{n,0}}+\beta\delta_{\omega_{n,0}}
\Sigma_1\tilde{g}\frac{\theta(s-s_c)}{s_c}
\eeq
of the propagator. In this representation of the Parisi matrices 
on the interval $[0,1]$, the replica indices are absent. Nonetheless, 
a well-defined formula\cite{parisi}
\beq
\frac1nTrAB=\tilde{a}\tilde{b}-\int_0^1 ds a(s)b(s)
\eeq
exists for taking the trace of a product of two ultrametric 
matrices $A$ and $B$, where $\tilde{a}$ and $\tilde{b}$ are the diagonal 
elements of $A$ and $B$ respectively and $a$ and $b$ are the 
corresponding off-diagonal elements in the continuous representation.
Let's consider here only a simple case of the weak RSB, 
$\beta \Sigma_1 \ll m^2$, assuming that $s_c \sim T$.
Substitution of Eqs. (\ref{propdia}) and (\ref{propoffdia})  
into Eq. (\ref{bm1}), and expanding over $\beta \Sigma_1 / m^2$
results in the following correction to the static conductivity
due to the replica symmetry breaking
\beq\label{sigcorr}
\delta\sigma^{\rm RSB}=\frac{2(e*)^2}{h} \eta \Sigma_1 \frac{1-s_c}{s_c}
\int_0^{\infty} \frac{x dx}{(x+m^2)^4} = 
\frac{4}{3} \frac{e^2}{h m^4}\eta (q_1-q_0)(1-s_c)
\eeq
Combining this with Eq.(\ref{sbos}), we obtain
\beq
\sigma^{\rm Tot}_{\rm bosons}=
\frac43\frac{e^2\eta}{h m^4}\left[ q_1+(q_0-q_1)s_c\right]
\eeq
as our total contribution for the bosonic conductivity.
If $s_c=1$, we recover our previous replica symmetric result. 
For the quantum $O(2)$ spin glass, however, $s_c\propto T$, 
and hence, the correction with $s_c$ vanishes at $T=0$.  
Setting $s_c=0$ requires that $q(s)=q_1$.  Hence, replica symmetry
breaking adds a simple benign constant to the conductivity which smoothly
crosses over to the replica symmetric result.

\section{Summary}

We have considered here two separate questions: 1) does the $O(2)$ vector
spin glass have a non-vanishing phase stiffness and 2) what is the role of 
replica symmetry breaking in the bosonic contribution to the conductivity.  
If the answer to the first question were yes, then the answer to the 
second would be irrelevant as the overall conductivity would be infinite.  
As we have demonstrated clearly, the spin glass order parameter does not 
provide a superconducting contribution to the conductivity 
at mean field and beyond.    
Our calculation of the phase stiffness seems to be the first based 
on a direct coupling of the vector potential to the spin glass order 
parameter which does not assume time translational invariance at the 
beginning. The physical mechanism underlying the vanishing of the spin 
stiffness appears to be the exploration of all configuration minima 
as a result of quantum tunneling.  In addition, we have found that replica 
symmetry breaking provides a small correction to
bosonic conductivity.  Hence, the bosonic metallic state we have found 
here is robust and represents a clear example of a metallic state in 
the presence of disorder in two dimensions.

\acknowledgements The calculations performed here were motivated by
a series of e-mails from S. Sachdev, C. Nayak, M. Fisher, and B. Halperin.
Our special thanks go to S. Sachdev who suggested that we reformulate our spin
glass problem and include the coupling to the vector potential and compute the 
conductivity exactly. We also thank E. Fradkin, G. Baym, and P. Fendley 
for useful discussions regarding analytical continuations. This work was 
funded by ACS Petroleum Research Fund.

\end{document}